\documentclass[pra, twocolumn, floatfix]{revtex4}

\usepackage{graphicx,textcomp}
\usepackage{dcolumn}
\usepackage{bm}
\usepackage[right]{eurosym}
\usepackage[ansinew]{inputenc}
\usepackage{amsmath}
\usepackage{amssymb}
\usepackage{color}
\usepackage{natbib}

\def\lssim{~\vcenter{\baselineskip 1pt    
\hbox{$<$}\hbox{$\sim$}}~}                
\def\gtsim{~\vcenter{\baselineskip 1pt    
\hbox{$>$}\hbox{$\sim$}}~}

\begin{document}

\title{Vibrational relaxation and dephasing of Rb$_2$ attached to helium nanodroplets}

\author{B. Grüner$^1$, M. Schlesinger$^2$, Ph. Heister$^1$\footnote[1]{Present address: Fakultät für Chemie, Technische Universität München}, W. T. Strunz$^2$, F. Stienkemeier$^1$, M. Mudrich$^1$}
\affiliation{$^1$Physikalisches Institut, Universit\"at Freiburg, 79104 Freiburg, Germany}
\affiliation{$^2$Institut f\"ur Theoretische Physik, Technische Universit\"at Dresden, 01062 Dresden, Germany}
\begin{abstract}
The vibrational wave-packet dynamics of diatomic rubidium molecules (Rb$_2$) in triplet states formed on the surface of superfluid helium nanodroplets is investigated both experimentally and theoretically. Detailed comparison of experimental femtosecond pump-probe spectra with dissipative quantum dynamics simulations reveals that vibrational relaxation is the main source of dephasing. The rate constant for vibrational relaxation in the first
excited triplet state $1^3\Sigma_g^+$ is found to be constant $\gamma\approx 0.5\,$ns$^{-1}$ for the lowest vibrational levels $v\lssim 15$ and to increase sharply when exciting to higher energies.\end{abstract}

\date{\today}

\maketitle

\section{Introduction}
Helium nanodroplet isolation (HENDI) is a well-established technique for isolating molecules and forming clusters at low temperature (0.38\,K) for spectroscopic studies~\cite{Toennies:2004}. Shifts and broadenings of spectral lines of molecules embedded in the helium droplets are small due to the weak dopant-host interactions as well as to the peculiar quantum properties of the superfluid helium nanodroplets~\cite{Grebenev:1998,Stienkemeier:2006}. Nevertheless, the details of the solute-solvent interactions that are at the origin of the observed line shapes are currently being studied with various approaches~\cite{Callegari:2000,Nauta4:2001,Dick:2001,Buenermann:2007,Lehnig:2009}. However, to date no time-resolved studies aiming at resolving the details of the interaction of excited molecules with helium nanodroplets have been performed.

Vibrational relaxation of molecules and molecular complexes embedded in helium nanodroplets has been studied by the group of R. Miller using high-resolution infrared spectroscopy and bolometric detection~\cite{Nauta:1999,Nauta2:2000,Nauta4:2001,Moore:2003}. It was found that systems having a large energy gap between the molecular vibration and the excitations of the helium (\textit{e.\,g.} HF ($v=1$)) couple very inefficiently to the helium environment, which leads to slow vibrational relaxation times $t\gtrsim 0.5\,$ms. The observed droplet-size dependent line shifts and broadenings point at coupling to surface excitations of the helium droplets (ripplons) being the main mechanism of relaxation~\cite{Moore:2003}. Using microwave spectroscopy, droplet-size dependent rotational relaxation times have been determined to be of order 1 -- 10\,ns~\cite{Callegari:2000}.

Recently, experiments have been performed that probe the dynamics of spin relaxation, dissociation, aggregation, exciplex formation, and vibration of molecules attached to helium nanodroplets~\cite{Koch:2008,Braun:2004,Przystawik:2008,Kornilov:2010,Droppelmann:2004,Mudrich:2008,Claas:2006,Claas:2007,Mudrich:2009}. These studies have in common that the helium nanodroplets act as a dissipative environment that decisively affects the outcome and the dynamics of the process of interest. Therefore, the dynamics of cooling, relaxation and dephasing induced by the ultracold bath of helium atoms attracts an increasing amount of attention both from experiment and theory.

The vibrational wave packet dynamics of alkali metal dimers attached to helium nanodroplets has been studied using the femtosecond pump-probe technique in a series of experiments in our group~\cite{Claas:2006,Claas:2007,Stienkemeier:2006,Mudrich:2009}. Alkali metal atoms and molecules represent a particular class of dopant particles due to their extremely weak binding to the surface of He droplets in bubble-like structures~\cite{Mayol:2005,Dalfovo:1994,Ancilotto:1995,Stienkemeier2:1995}. In particular, pump-probe measurements with K$_2$ diatomic molecules in singlet states attached to helium droplets reveal a significant impact of the helium environment on the vibrational dynamics, suggesting the manifestation of a Landau critical velocity for the vibrational motion of K$_2$ on the surface of superfluid helium nanodroplets~\cite{Claas:2006,Schlesinger:2010}.

In this work we present the detailed analysis of pump-probe measurements of the vibrational wave packet dynamics of Rb$_2$ molecules in triplet states attached to helium nanodroplets with regard to relaxation and dephasing induced by the helium environment. This system is particularly well-suited for a quantitative study of the molecule-helium droplet interaction due to the precise knowledge of the spectra and dynamics of gas-phase Rb$_2$ molecules and due to the weak molecule-helium interactions which allow for an accurate theoretical description.

The long-lasting wave packet oscillations that we observe up to delay times $t \gtsim 1.5\,$ns suggested that the Rb$_2$ molecules desorb off the helium droplets on a short time scale $t \lssim 10$\,ps and continue to vibrate freely in the gas-phase~\cite{Mudrich:2009}. This assumption was backed by the good agreement between the measured and theoretically predicted vibrational frequencies. Besides, the measurement of beam depletion using a separate detector was interpreted as clear evidence that excited Rb$_2$ molecules desorb off the droplets on the time scale of the flight time from the laser interaction region to the detector ($\sim 1\,$ms). Furthermore, earlier measurements with K$_2$ dimers as well as theoretical simulations on K atoms attached to helium nanodroplets indicated desorption times in ranging 3 -- 8\,ps and 10 --  30\,ps, respectively~\cite{Claas:2006,Takayanagi:2004}.

A more detailed inspection of our data reveals, however, that the Rb$_2$ molecules are subject to continuous vibrational relaxation due to the constant coupling to the bath of helium atoms on the time scale of the pump-probe measurements. The experimental signature of the coupling of vibrating Rb$_2$ to the helium is the decreasing contrast of wave packet oscillation signals as well as changing amplitudes of individual Fourier frequency components due to the redistribution of populations of vibrational states. In particular, the pronounced dependence of the dephasing time on the quantum number $v$ of excited vibrational levels points at system-bath couplings being active. This observation is in line with earlier measurements of the fluorescence emissions of Na$_2$ molecules in triplet states, which indicated vibrational relaxation in the excited electronic state to take place on the time scale of the life time of the excited state due to spontaneous emission ($\sim 10$\,ns)~\cite{Bruehl:2001}. Recently, it was observed that desorption upon electronic excitation may even be completely inhibited in the case of Rb atoms excited in a particular laser wave length range~\cite{Auboeck:2008}.

Pioneering experiments on the vibrational dephasing and relaxation of molecules ($I_2$) exposed to collisions with rare gas atoms at high density were performed by the Zewail group~\cite{Liu:1996}, motivating theoretical studies by Engel, Meier~\textit{et al.}~\cite{Ermoshin:1999,Meier:2004}.
More recently, dephasing times as well as relaxation rates have been studied extensively by the groups of Apkarian and Schwentner by means of femtosecond spectroscopy of the vibrational dynamics of halide molecules isolated in cryogenic rare-gas matrices~\cite{Karavitis:2003,Karavitis:2004,Kiviniemi:2005,Guehr:2004,Fushitani:2005,Guehr:2007}. Seminal work on time-resolved measurements of the dissipative fluid dynamics in bulk He-II has been performed using femtosecond pump-probe spectroscopy of triplet He$_2^*$ excimers created inside He-II~\cite{Benderskii:2002}. Due to the strong coupling of the highly excited He$_2^*$ to the surrounding He which forms an extended bubble around He$_2^*$ the dynamics is fully damped after one period of motion.

Theoretical studies on collisional quenching of rotations and vibrations of alkali dimers and other small molecules by helium atoms at low temperatures have recently been stimulated by the prospects of creating samples of cold molecules using buffer-gas cooling as well as sympathetic cooling with ultracold atoms as a cooling agent~\cite{Bodo:2006,BodoReview:2006,Caruso:2010}. In the system Li$_2$+He, for instance, the quenching rate constants in the  approximation of vanishing temperature are predicted to increase by about one order of magnitude with increasing vibrational levels $v=0 - 10$~\cite{Bodo:2006}.

In the case of weak couplings, which applies to our system, the concept of perturbations of the vibrational levels of the molecules by fluctuations in the bath modes is well-established. It leads to vibrational energy relaxation along with the decay of vibrational coherences.
 There may well be additional pure dephasing mechanisms that are expected to vanish in the low
temperature limit \cite{Zewail:1979,Karavitis:2005}.
In the related energy gap picture the populations of individual vibrational energy levels $v$ relax stepwise to the next lower vibrational energy levels $v-1$~\cite{Nitzan:1975}. Here, the relaxation rate for an isolated vibrational level increases with the vibrational quantum number $v$~\cite{Englman:1979,Kiviniemi:2005,Karavitis:2005}. The evolution of coherences and populations of vibrational levels in the weak-coupling limit is often being modeled with the master equation description, obtained from the anharmonic molecular oscillator coupled to a harmonic bath. Various coupling terms are used for describing different interaction mechanisms~\cite{Bader:1996,Foeldi:2003,Gershgoren:2003,Kiviniemi:2005,Karavitis:2005}. \\
Following established practice in the chemical literature,
in this contribution we use ``dephasing'' to describe the general mechanism of loss of coherence between quantum states.
Note that in other fields, this process is preferably referred to as ``decoherence'' while ``dephasing'' is then used
to describe that special occurrence of decoherence, where no dissipation is involved.

\section{Vibrational wave packet dynamics} \label{sec:wavepacketdynamics}
The experimental arrangement used for recording femtosecond pump-probe photoionization transients is identical to the one described previously~\cite{Mudrich:2009}. In short, a continuous beam of helium nanodroplets of the size of about 8000 He atoms is produced by expanding high-purity $^4$He gas out of a cold nozzle (T$\approx 17\,$K, diameter $d=5\,\mu$m) at high pressure (p$\approx$50\,bar). The helium droplets are doped with two Rb atoms on average per droplet by passing through a pick-up cell that contains Rb vapor at a pressure $p_{Rb}\approx 2\times 10^{-4}\,$mbar. Alkali atoms and molecules are peculiar dopants in that they reside in
bubble-like structures on the surface of He droplets. Upon formation of a Rb$_2$ diatomic molecule, the binding energy is dissipated by evaporation of helium atoms from the droplets and occasionally by desorption off the droplets of the newly formed molecule itself. This leads to an enrichment of droplet-bound Rb$_2$ molecules in weakly bound triplet states.

Further downstream, the doped He droplet beam intersects the laser beam inside the detection volume of a commercial quadrupole mass spectrometer. Due to the limited mass range, only bare Rb$_2^+$ photoions are detected mass-selectively. The laser beam consists of pairs of identical pulses produced by a commercial mode-locked Ti:sapphire laser and a Mach-Zehnder interferometer to adjust the time delay between the pulses. The pulses have a duration of $\approx 160\,$fs and a spectral bandwidth at half maximum of $\Delta\omega_{las}\approx 80\,$cm$^{-1}$ and peak pulse intensity $\sim 4\,$GW/cm$^2$.

\begin{figure}
	\centering
        \includegraphics[width=1.0\columnwidth]{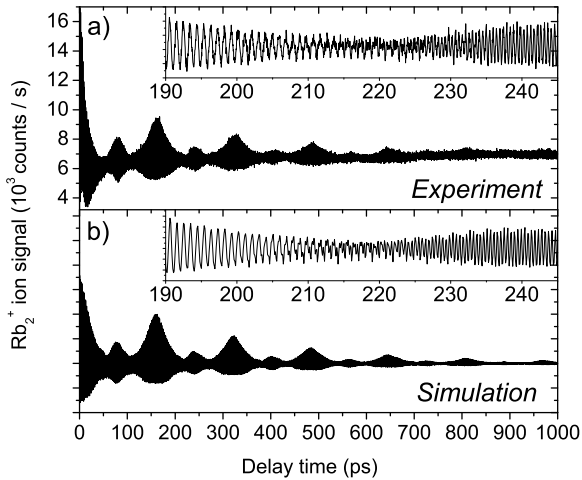}
	\caption{Experimental (a) and simulated (b) pump-probe transients of Rb$_2$ formed on helium nanodroplets recorded at the laser wave length $\lambda = 1006\,$nm.}
	\label{fig:PPsignal}
\end{figure}
A pronounced oscillatory photoionization signal is observed in the pump-probe transients for laser wave lengths in the range $\lambda=960$\,nm -- $1032$\,nm. A typical measured pump-probe transient recorded at $\lambda=1006\,$nm is depicted in Fig.~\ref{fig:PPsignal} (a). On the time scale of picoseconds, the transient signal is modulated by wave packet (WP) oscillations with a period $T_{\Sigma g}\approx 0.95\,$ps (see inset). In addition, this oscillation is amplitude- and frequency-modulated due to dispersion and subsequent revivals of the WP motion in an anharmonic potential. On the long time scale of hundreds of ps, the contrast of WP oscillations degrades monotonically and eventually vanishes at delay times $\gtrsim 1\,$ns.
In this paper we argue that this slow decay of contrast is due to dephasing induced by the coupling of the Rb$_2$ molecules to the helium droplets. Other sources of dephasing are conceivable: gas phase collision with evaporated gas and clusters of helium or dephasing due to the influence of rotations. However,
 an estimate of the Rb$_2$-He gas phase collision rate gives a value far too small to account for the observed data.
As for the influence of rotations, it is clear that an initial thermal population of rotational levels leads
to a similar decay of signal contrast as observed in the experiment \cite{Schlesinger:2008}. Again, in our case a detailed analysis
shows that in order to account for the observed decay rates, unphysically large temperatures would have to be assumed.
Moreover, the observed functional dependence of decay rates on the laser wavelength cannot easily be explained.
Taking into account all these findings, our picture of vibrational damping of the dimer through the interaction
with the helium droplet allows for the most consistent explanation of all observed phenomena.

\subsection{Free gas phase dynamics}\label{sec:GasephaseDynamics}
\begin{figure}
	\centering
		\includegraphics[width=0.95\columnwidth]{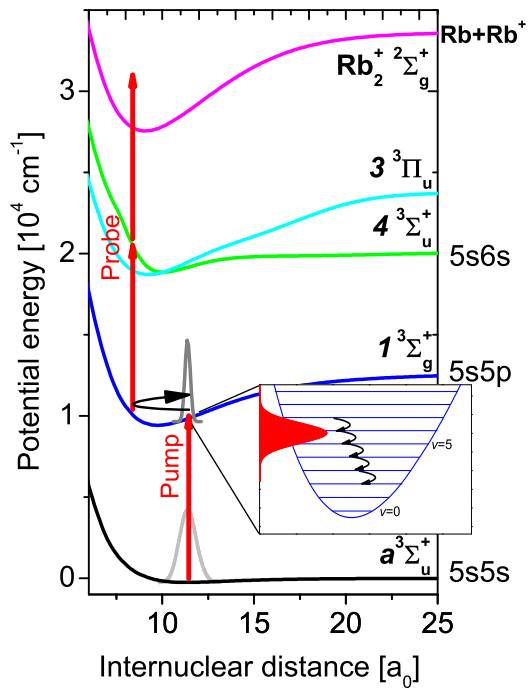}
	\caption{(Color online) Selected triplet potential-energy curves of
neutral Rb$_2$ and of the Rb$_2^+$ ionic ground state relevant to the
present study. The arrows indicate the creation of a vibrational wave packet in the
first-excited state followed by resonant two-photon ionization.}
	\label{fig:Potentials}
\end{figure}
Let us first briefly review the fundamental aspects of pump-probe spectroscopy of diatomic molecules isolated in the gas-phase. A first pump pulse excites a coherent superposition of vibrational states (WP) in an  excited electronic state.
After some time delay a second probe pulse projects the WP to a final ionic state which is detected as a function of time delay between the two pulses. Fig.~\ref{fig:Potentials} shows the relevant potential energy curves of the $\mathrm{Rb}_2$ molecule in the triplet manifold.
The straight arrows symbolize the pump and probe pulse excitation pathways.

Due to the cold helium environment, at time $t=0$ only the vibrational
ground state on the lowest triplet state $a {^3 \Sigma^+_u}$ is occupied.
Detailed experimental and theoretical
analysis reveals that rotational degrees of freedom
need not be taken into account (see also Sec.~\ref{PumpProbeSpectra}).
As outlined in Sec.~\ref{sec:simulation}, we fully solve the time-dependent Schr\"odinger equation
involving all relevant potential energy surfaces. Here, we want to point out that the
main effect of the pump pulse is to generate a coherent superposition of vibrational
eigenstates $|v\rangle$ on the  excited electronic state surface
$\left( 1 \right)^{3}\Sigma^{+}_{g}$. More specifically, the created WP
can be written as $\left|\psi_e (t=0) \right\rangle = \sum_{v =0}^N c_v \left|v\right\rangle$,
where $c_v$ denotes complex expansion coefficients and $v=0,1,2,\dots$ is the vibrational quantum number.
Coherent WPs will also be created in higher lying electronic
states or, by resonant impulsive stimulated Raman scattering (RISRS) in the triplet ground state, with a significantly smaller amplitude, though~\cite{Chesnoy:1988}. Thus, all linear and nonlinear processes are fully taken into account in our simulations.
The respective populations of vibrational levels $|c_v|^2$ depend
on the pump pulse parameters and on the Franck-Condon factors of the excitation transition. The pump pulse wave length determines the central vibrational level, while the pulse width, the pulse energy and the Franck-Condon factors determine the number and the relative populations of vibrational levels. The created WP in the $\left(1\right)^{3}\Sigma^{+}_{g}$-state propagates in the region between the classical inner and outer turning points. Direct integration of the Schr\"odinger equation $i \hbar \partial_t\left|\psi_e\right\rangle =
H_e \left|\psi_{e}\right\rangle$ yields
\begin{equation}\label{eq:9}
\left|\psi_{e} \left( t \right) \right \rangle = \sum\limits_{v =0}^N c_v e^{ -\frac{\mathrm{i}E_v t}{\hbar} } \left|v\right\rangle,
\end{equation}
where $E_v$ is the energy of the $v$-th vibrational level.
As discussed below, the measured signal allows to extract information about the density matrix of vibrational states $\rho_e(t)$.
For the isolated dimer, $\rho_e(t)= | \psi_e(t) \rangle \langle \psi_e(t) |$ describes a pure state at all times.
Its time evolution is given by
\begin{equation}\label{eq:5}
\rho_e \left(t\right)= \sum \limits_{v,v'} c_v c_{v'}^*
e^{-\frac{i \left(E_v - E_{v'} \right)t  }{\hbar}} \left|v\right\rangle \left\langle v' \right|.
\end{equation}
The diagonal elements of the density matrix $\rho_{vv}=|c_v|^2$ represent the populations that
are constant in time, while the off-diagonal elements $\rho_{v v' }(t) \equiv \langle v  | \rho_e(t) | v' \rangle $ with $v  \neq v'$ oscillate
 with Bohr frequencies $ \omega_{v v'}= (E_v -E_{v'}) / \hbar$ and represent the coherences between the vibrational eigenstates $| v \rangle$, $| v' \rangle$.

The probe pulse produces photoions through a resonant 2-photon-transition from the excited state to the ionic ground state $^{2}\Sigma^{+}_{g}$ of Rb$_2^+$. Transitions preferably take place when the WP is located around a well-defined transition region, where the transition dipole matrix element is maximal (Franck-Condon window).
Even though our simulations are numerically exact, it is instructive
to consider the perturbative dependence of the ion signal on the WP density matrix \cite{Gruebele:1990, Engel:1991, Seel:1991,Seel:1991a},
\begin{equation}\label{eq:4}
S(t)=\sum_{v v'  }A_{v v' } \rho_{v v' }(t).
\end{equation}
The coefficients $A_{v v' } $ in Eq.~(\ref{eq:4}) contain
products of transition moments and field parameters and provide information about the
vibrational populations in the final state.
Through the dependence on the density matrix, the signal $S$ is composed of beat frequencies $\omega_{v v'}$ between all pairs of energy levels that contribute to the WP.
The most prominent oscillation in the signal originates from components $\omega_{v v+1}$ and reflects the circulation of the WP on the potential energy surface. From the Fourier spectra of the signal information about higher-order frequency components $\omega_{v v+\Delta v}$ with $\Delta v > 1$ can be extracted. In this way it is possible to gain information about the density matrix from the measured ion signal.

In a harmonic potential with energy levels $E_n = \hbar \omega_e n$, where $n=0,1,2,\dots$ denotes the number of eigenstate $\left|n\right\rangle$, Eq.~(\ref{eq:9}) yields a periodic oscillation with classical period
$T_c=h/ \Delta E = 2 \pi/ \omega_e$, where $\Delta E $ denotes the constant energy spacing between adjacent levels $n$.
The signal $S$ in Eq.~(\ref{eq:4}) features a periodic oscillation with period $T_c$.
In the anharmonic Morse potential with energy spectrum $E_v = \hbar \omega_e(v -  x_e v^2)$, however, initially well-localized WPs spread out due to dispersion on the characteristic time scale $T_{\mathrm{disp}}=\hbar \hat{\omega} /(\omega_e x_e \Delta E_{\mathrm{pump}})$ \cite{Averbukh:1989,Vetchinkin:1994,Guehr:2007}.
Here, $ x_e$ is the anharmonicity constant,
$\hat{\omega}$ denotes the central vibrational frequency of the WP and $\Delta E_{\mathrm{pump}}$ is the spectral energy width of the pump laser pulse.
Considering the parameters of our experiment we obtain $T_\mathrm{\mathrm{disp}}\sim 100\,\mathrm{ps}$.
 Around $t=T_{\mathrm{disp}}$, all contributions
 on the right-hand side of Eq. (\ref{eq:4}) appear uncorrelated, which means that the oscillatory signal collapses~\cite{Scully:1997}.
 Therefore, dispersion of the WP leads to a decay of the pump-probe signal contrast.
 Note, however, that due to dispersion neither populations $\rho_{v v} $ nor the absolute values of coherences $\left|\rho_{ v v'}\right|$, $v \neq v'$, change in time.

 In the Morse potential, a revival of the initial WP
 takes place at certain times, i.\,e.\,the original phase correlation in the WP
 is restored and the WP partly or fully revives.
At the full revival time, the signal amplitude ideally reaches its initial height, which underlines that
coherence is preserved.
Full revivals occur at times $t=k\times T_{\mathrm{rev}}/2$, where $ T_{\mathrm{rev}}= 2\pi / (\omega_e x_e) $, when all vibrational eigenstates have accumulated a phase of $2 \pi k$ with $k=1,2,3,\dots$. At fractions of the revival time, $t=p/q\times T_\mathrm{rev}$ where $p/q$ is an irreducible fraction of integers, the WP consists of a superposition of $q$ copies of the original WP (fractional revivals)~\cite{Averbukh:1996,Vrakking:1996}. For instance, at half-period revivals ($p/q=1/2$) the initial well-localized WP evolves into a highly quantum mechanical state that consists of two counter-propagating partial WPs that interfere with each other when colliding.
In the electronic state of relevance for the present analysis, $1^3\Sigma_g^+$ of Rb$_2$, which perfectly matches the shape of the Morse potential in the accessible range of $v$-states, the first full revival time is $T_\mathrm{\mathrm{rev}}/2\approx 160\,\mathrm{ps}$~\cite{Mudrich:2009}.

\subsection{Wave packet dynamics with dissipation}\label{sec:DynamicsWithDissipation}

The previous discussion was devoted to isolated vibrating diatomic molecules in the gas-phase. Let us now consider Rb$_2$ molecules (M) coupled to the dissipative environment realized by helium nanodroplets (HND), to which the molecules are attached.


A Rb$_2$ molecule attached to a HND is a closed but complicated system which can be described by the Hamiltonian $H=H_{\mathrm{M}}+H_{\mathrm{HND}}+H_{\mathrm{M \leftrightarrow HND}}$. $H_{\mathrm{M}}$ denotes the isolated molecule as discussed before, $H_{\mathrm{HND}}$ is the Hamiltonian for the pure helium nanodroplet and $H_{\mathrm{M \leftrightarrow HND}}$ contains the interaction between the two. As we are only interested in the dynamics of the molecule we call this our "system" and the helium nanodroplet our "bath".
In our experiments the WP dynamics of the coupled Rb$_2$ molecule is mostly very similar to that in the gas phase, which means that we see the same fast oscillations, WP dispersion with time constant $T_\mathrm{disp}$,
and (fractional) revivals at times $T_\mathrm{rev}$. However, on the long time scale of the experiment (nanoseconds)
the oscillatory signal exponentially decays due to slow system dephasing. Thus, a description in terms of a weak system-environment coupling is justified.

In the experiment, we observe a decay of the revival amplitudes with a rate $\gamma_{D}$. As we will show,
this decay is related to the environment-induced dephasing of the WP due to dissipation.
Dephasing can also be caused by a process which only affects the off-diagonal elements of the density matrix (no dissipation). This process is referred to as "pure dephasing". 
For the well-known two-level system, the overall dephasing time constant $T_2$ is related to relaxation ($T_1$) and pure dephasing without dissipation ($T_*$) by $1/T_2 = 1/(2\,T_1)+1/T_*$.
 For multi-level oscillators, which we consider here, the relation between dephasing and
dissipation is more subtle. No simple general expression connecting the corresponding
time scales exists: depending on the shape of the WP, dephasing may take
place on a much shorter time scale \cite{Joos:1985,Zurek:1991,Giulini:1996,Tegmark:2001}. We recall that it
is important to distinguish between
contrast decay due to dephasing -- which is an irreversible process --
and the reversible drop of the observed oscillation amplitude due to dispersion
 in an anharmonic potential.

The dissipative vibrational dynamics is described using the framework of
Markovian master equations. At this stage, we do not aim at deriving such an
equation from a microscopic Hamiltonian, which would require detailed knowledge of the
helium ``bath'' and interaction Hamiltonians, $H_\text{HND}$ and $H_{\mathrm{M \leftrightarrow HND}}$, respectively. Instead, we choose a well-established
Markovian quantum optical master equation \cite{Breuer:2002} for a weakly
coupled environment. The density operator $\hat{\rho}(t)$ of the (reduced) system that describes dissipation in near-harmonic systems at zero temperature is given by
\begin{equation}\label{eq:3}
	\partial_t \hat{\rho} =\frac{1}{ i \hbar} \left[\hat{H}_M,\hat{\rho}\right]
       + \underbrace{\sum\limits_j\left( \hat{L}_j \hat{\rho} \hat{L}_j^\dagger - \frac{1}{2} \{  \hat{L}_j^\dagger \hat{L}_j, \hat{\rho} \} \right) }_{\text{coupling to bath}}.
\end{equation}
This equation is of Lindblad form \cite{Lindblad:1975}.
To describe dissipation, we use $L_i=\sqrt{\gamma_i}\hat{a}_i$, where $\hat{a}_i$ is
the usual quantum mechanical ladder operator,
defined through the harmonically approximated potential energy curve $i$.
This Lindblad operator induces vibrational dissipation on the time scale $1/\gamma_i$.
More specifically, independently of the initial conditions, to good approximation,
the mean energy of the excited WP decreases exponentially with a respective rate $\gamma_i$.

The relaxation rate constants $\gamma_i$ in Eq.~(\ref{eq:3})
are taken as fit parameters to match the experimental data.
The chosen dissipative Lindblad operator 
also affects the off-diagonal elements of the
density matrix. The latter decay with time due to dephasing which implies a transition from
an initially pure state to a state mixture. For $ t \rightarrow \infty $, only the ground
vibrational state is occupied, which is, of course, a pure state again.
For the master equation (\ref{eq:3}) it is well known that localized WPs
are ``robust'' in the sense that they suffer only little dephasing. In contrast, two such WPs separated by a (dimensionless) distance $D$ in phase-space loose their
coherence with an accelerated rate $D^2\gamma$~\cite{Caldeira:1985,Waals:1985,Brune:1992,Braun:2001}.

\begin{figure}
	\centering
		\includegraphics[width=1.05\columnwidth]{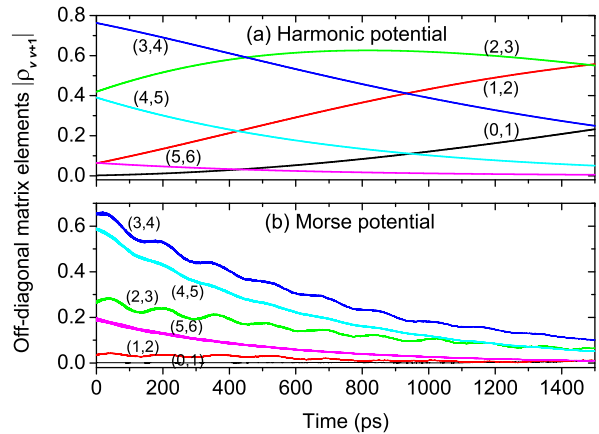}
	\caption{(Color online) First-order coherences (absolute values of the off-diagonal elements of the density matrix, $\rho_{v\,v+1}$), $v=0$--$6$, for vibrational wave packet dynamics in the $1^3\Sigma_g^+$-state potential of Rb$_2$ (b) in comparison with the harmonically approximated potential (a).}
	\label{fig:HOvsMorse}
\end{figure}
This effect is visualized in Fig.~\ref{fig:HOvsMorse}, in which the time evolution of the first-order coherences
$|\hat{\rho}_{v v+1}|$ of the dynamics in the Morse-type potential of the $1^3\Sigma_g^+$-state of Rb$_2$ (b) is compared with the dynamics in the harmonically approximated potential (a). The shown initial distribution of coherences is obtained when exciting a WP formed of vibrations $v~\approx 0$--$6$ as observed in the experiment at a laser wave length $\lambda =1025\,$nm. The matrix elements $\hat{\rho}_{v v'}$ are computed by solving Eq.~(\ref{eq:3}) numerically and by projecting onto the eigenstates $\left|v\right\rangle$ at every time step. In the case of a harmonic potential (Fig.~\ref{fig:HOvsMorse} (a)), beats between low-lying levels $n\lssim 3$ initially rise on the shown time scale, whereas the high-lying level beats monotonically fall to zero. This is due to the vibrational redistribution from the high-lying levels into the lower-lying ones, which evidently not only redistributes population but also transfers coherence from the high-lying to the lower-lying levels.

In the case of the Morse-type potential (Fig.~\ref{fig:HOvsMorse} (b)), this tendency is much less pronounced. All the shown coherence terms drop down with slightly varying decay rates. The slower decay of the low-level beats $(1,2)$ and $(2,3)$ in contrast to the faster decay of $(3,4)$ and $(4,5)$ is partly reminiscent of the vibrational redistribution mentioned before. In addition, higher excited levels decay faster. Thus, for the Morse oscillator the overall dephasing appears to be accelerated with respect to the harmonic oscillator. Moreover, beats $(2,3)$ and $(3,4)$ in Fig.~\ref{fig:HOvsMorse} (b) are periodically modulated with period $T_{\mathrm{rev}}/2$. In particular, a slow loss of coherence or even a momentary increase of coherence is apparent at times close to the full vibrational recurrences when the WP  is well-localized again. Around the half-period fractional revivals, when the WP splits into partial WPs that are maximally delocalized (large $D$), however, dephasing is fastest.

\subsection{Numerical simulation}\label{sec:simulation}
In order to reproduce the ion yield in the gas phase,
we calculate the final state probability after the interaction with the laser field.
For the isolated dimer, we here follow the approach of \cite{Vivie-Riedle:1996} and
fully numerically solve the time dependent Schr\"odinger equation
\begin{equation}\label{eq:1}
\partial_t | \Psi(t)\rangle = -\frac{i}{\hbar}H_{\mathrm{M}} | \Psi (t) \rangle
\end{equation}
for the full state vector $\Psi=(\psi_g,\psi_e,\dots)$.
The Hamilton operator $H_{\mathrm{M}}$ now also contains the field interaction with the molecule.
In particular, we take into account that the final state consists of the bound
ion plus an ejected electron with energy $E$.
Following the approach of \cite{Vivie-Riedle:1995}, we use a discretization of the electronic continuum.
We determine the final state probability $|\psi_f(E,\tau)|^2$
for different pump-probe delays $\tau$ and electronic energies $E$.
Adding contributions with different energies $E$, we obtain a signal $S(\tau)$, which is
proportional to the gas phase ion yield. For the isolated molecule, Eq.~(\ref{eq:1}) also directly allows to obtain
the density operator through $\hat{\rho}(t)= |\Psi(t) \rangle \langle \Psi(t)|$.
Potential energy surfaces
and transition dipole moments were provided by O. Dulieu \cite{Beuc:2007}.

For our phenomenological description of the helium influence on the dimer dynamics,
we switch to the density matrix description, Eq.~(\ref{eq:3}).
Our aim is to ascribe certain damping parameter values $\gamma_i(\lambda)$ to the measured pump-probe signal at wave lengths $\lambda$.
From a numerical point of view,
the evolution of the density matrix can become very costly,
in particular, if one considers many potential energy surfaces and/or many
vibrational states. We therefore return to an equation for the state vector, \textit{i.\,e.}
to a Schr\"odinger-type equation including relaxation (and thus, dephasing).
The density matrix of the master equation (\ref{eq:3}) is recovered from the stochastic Schr\"odinger equation on average, $\hat{\rho}(t)=\overline{|\Psi(t) \rangle \langle \Psi(t)|}$ \cite{Gisin:1992}.
In practice, one has to determine many realizations of state vectors  $\Psi_{\text{SSE},i}$, which can then be used
to extract the density matrix (Fig.~\ref{fig:HOvsMorse}), coordinate, momentum, energy expectation values 
, or the final state probability (Fig.~\ref{fig:PPsignal} (b)) to compare with the experiment.

\section{Pump-Probe Spectra}\label{PumpProbeSpectra}
Upon laser excitation of Rb$_2$ molecules formed on helium nanodroplets at wave lengths in the range $\lambda=960$\,nm -- $1032$\,nm, coherent vibrational WPs are created in the first excited triplet state $1^3\Sigma_g^{+}$ as well as in the lowest triplet state $a^3\Sigma_u^{+}$ by RISRS with varying relative intensity. Around $\lambda=1010\,$nm, the pump-probe signal as shown in Fig.~\ref{fig:PPsignal} (a) is dominated by WP motion in the $1^3\Sigma_g^+$-state. The amplitude modulation results from dispersion of the WPs and half-period as well as full recurrences are observed with high contrast at revival times $T_{\mathrm{rev}}/4\approx 80\,$ps and $T_{\mathrm{rev}}/2\approx 160\,$ps, respectively. The nearly exponential decrease of the signal contrast is attributed to relaxation-induced dephasing and will be investigated in detail in the following. The simulated transient ($\gamma_{\Sigma g}=0.45\,$ns$^{-1}$), depicted in Fig.~\ref{fig:PPsignal} (b), nicely reproduces both the vibrational recurrences as well as the overall damping due to vibrational dephasing.


At short delay times $t\lssim 50\,$ps we observe slight deviations between the simulated and experimental transient signals, which
we attribute to the influence of rotations, as seen previously for iodine molecules~\cite{Gruebele:1993}.
As mentioned in the introduction, a dephasing influence of rotations on pump-probe spectra is
conceivable~\cite{Schlesinger:2008}, yet requires unphysically high temperatures $\gtrsim 10\,K$ in our case.

Gas phase simulations of the vibrational WP dynamics including free rotations in the Rb$_2$ system feature notable rotational recurrences at half and full rotational periods $T_{\mathrm{rot}}/2\approx 575\,$ps and $T_{\mathrm{rot}}\approx 1150\,$ps, respectively.
However, no such signals are observed in the experimental data. Due to the surface position of the Rb$_2$ molecules presumably with the molecular axis being oriented parallel to the surface~\cite{Bovino:2009}, we expect the rotation to decompose into weakly perturbed in-plane rotation and strongly hindered out-of-plane rotation which more likely resembles a pendular motion. The latter may efficiently couple to surface modes of the droplets causing fast relaxation. 
Couplings between vibration, rotation, and libration may therefore
induce intricate relaxation dynamics, which, however,
would require more expanded simulations that lie
beyond the scope of the present work

\begin{figure}
	\centering
		\includegraphics[width=0.95\columnwidth]{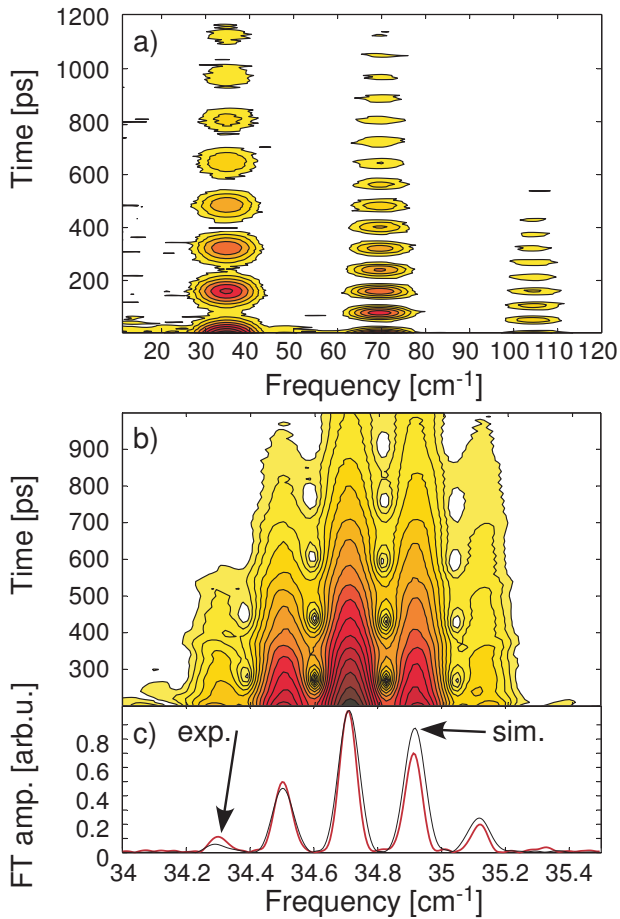}
	\caption{(Color online) Sliding window Fourier spectra (spectrograms) of the pump-probe transients recorded at $\lambda = 1006\,$nm. The width of the time window is $5\,$ps in (a) and $400\,$\,ps in (b). (c) displays the power spectrum of the integral experimental and theoretical transients.}
	\label{fig:Spectrogram1006nm}
\end{figure}
The experimental pump-probe signal of Fig.~\ref{fig:PPsignal} (a) is analyzed by Fourier transforming the time trace inside a time window of a width of 5 ps and a Gaussian apodization function with full width at half maximum (FWHM) of $2.6\,$ps that slides across the data (wavelet or spectrogram analysis). This type of analysis provides information about the frequency components that make up the WPs without losing the complete dynamical information. The result for $\lambda=1006\,$nm
is displayed in Fig.~\ref{fig:Spectrogram1006nm} (a). In this representation, the individual WP oscillations are no longer resolved, but the full, half-period and even one third-period revivals are clearly visible and can be attributed to frequency-beats between vibrational states separated by $\Delta v=1$, $2$, and $3$ vibrational quanta, respectively~\cite{Vrakking:1996}. The fact that the WP recurrences are seen with such an extraordinarily high contrast even at long delay times is a consequence of the shape of the $1^3\Sigma_g^+$-potential that nearly perfectly matches that of the Morse potential in combination with weak system-bath couplings~\cite{Mudrich:2009}.

Fig.~\ref{fig:Spectrogram1006nm} (b) displays a magnified view of the spectrogram of the same data when using a time window of width $400\,$ps and an apodization function with FWHM $209\,$ps in the spectral range $\nu=34$--$35.5\,$cm$^{-1}$. In this representation of the data, the frequency resolution is comparable to the one obtained by transforming the integral data set (Fig.~\ref{fig:Spectrogram1006nm} (c)) while still retaining the dynamics on the long time-scale. The individual frequency components reflect beats between coherently excited adjacent vibrational states that are unequally spaced due to the anharmonicity of the potential. By comparing to the Fourier spectrum of the simulated data  in Fig.~\ref{fig:Spectrogram1006nm} (c) we conclude that the WPs excited at $\lambda=1006\,$nm are composed of vibrational states $v=6$--$11$ with relative amplitudes determined by the spectral intensity profile of the fs laser.

\subsection{Analysis of dephasing dynamics}
\begin{figure}
	\centering
		\includegraphics[width=0.95\columnwidth]{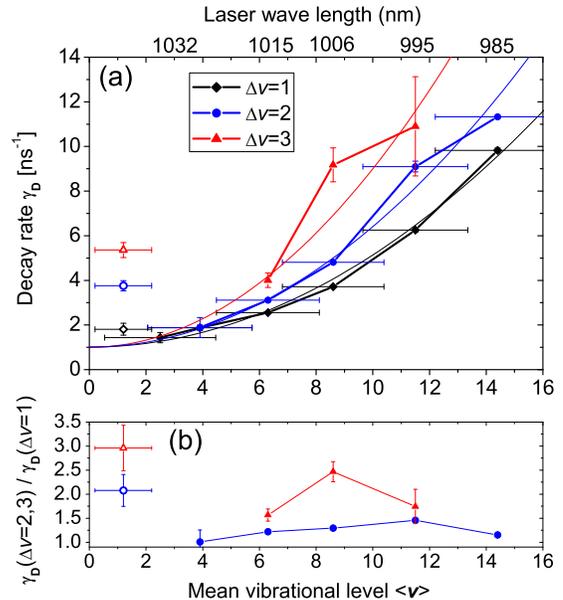}
	\caption{(Color online) Exponential decay time constants obtained by fitting the maxima of the full ($\Delta v=1$) as well as fractional ($\Delta v=2,3$) revivals plotted against the average vibrational quantum number of vibrational states initially populated by the pump pulse $\left\langle v\right\rangle$. The filled symbols refer to the wave packet dynamics in the excited $1^3\Sigma_g^+$-state, the open symbols refer to the $a$-state.}
	\label{fig:DecayConstants}
\end{figure}
In a first attempt to analyze the loss of contrast of the coherent WP oscillation signal the envelopes of the beat signals corresponding to $\Delta v=1,\,2,\,3$ are extracted from the spectrograms of the type shown in Fig.~\ref{fig:Spectrogram1006nm} (a) at various laser wave lengths. Due to the low spectral resolution the few vibrational beats that are simultaneously excited according to the laser band width are not resolved. Thus, each envelope trace is composed of the sum of individual beats excited around the central laser wave length. The envelope traces are fitted to an exponential decay function $\propto\exp (-\gamma_\mathrm{D} t)$ to infer the characteristic decay rate $\gamma_\mathrm{D}$.

The fit results for $\gamma_{\mathrm{D}}$ are depicted in Fig.~\ref{fig:DecayConstants} as a function of the mean vibrational quantum number $\left\langle v\right\rangle$ determined by the central laser wave length. The horizontal error bars reflect the width of the distribution of excited vibrational levels due to the laser band width, the vertical error bars depict the fit errors. Strikingly, the decay of contrast is strongly dependent on the level of vibrational excitation and features rapidly increasing decay rates $\gamma_\mathrm{D}$ with increasing $v$. The solid lines in Fig.~\ref{fig:DecayConstants} (a) represent model curves obtained by fitting quadratic functions to the data.

Similar behavior was observed in time-resolved coherent anti-Stokes Raman-scattering measurements of the WP dynamics of molecular iodine $I_2$ in the groundstate isolated in rare-gas cryo-matrices~\cite{Karavitis:2003,Karavitis:2004,Kiviniemi:2005}. There, the transition from a linear $v$-dependence of $\gamma$ to a quadratic dependence with increasing temperature of the matrix was observed. Linear $v$-dependence at low temperatures was interpreted in terms of dephasing induced only by vibrational energy relaxation whereas at higher matrix temperatures pure elastic dephasing also contributed.

Coherences between vibrational states spaced by $\Delta v=n>1$ are clearly subject to enhanced decay (Fig.~\ref{fig:DecayConstants} (a)). The experimental relative decay times amount to $\gamma_\mathrm{D}^{\Delta v=2}/\gamma_\mathrm{D}^{\Delta v=1}\approx 1.3$ and $\gamma_\mathrm{D}^{\Delta v=3}/\gamma_\mathrm{D}^{\Delta v=1}\approx 2$ as shown in Fig.~\ref{fig:DecayConstants} (b). The scaling behavior of these multi-order decay rates $\gamma_\mathrm{D}^{\Delta v=n}$ with the order $n$ of the beat has been discussed in the context of different mechanisms of pure dephasing~\cite{Gershgoren:2003}. Depending on the collision model considered in that study, a scaling behavior ranging from zeroth to second order with $n$ was expected. Our approach would require the inclusion of pure dephasing terms to account for $\Delta v$-dependent decay times whereas
in the current model, where dephasing is solely induced by dissipation, the revival decay times are found to be
independent of $\Delta v$.

The open symbols in Fig.~\ref{fig:DecayConstants} depict decay rate constants $\gamma_\mathrm{D}^a$ for the WP dynamics in the lowest triplet $a^3\Sigma_u^+$-state.
While $\gamma_\mathrm{D}^a$ of the first order coherence ($\Delta v=1$) is similar to those of the excited $1^3\Sigma_g^+$-state for small $\left\langle v\right\rangle$, $\gamma_\mathrm{D}^a$ for the higher order coherences are significantly higher than for the $1^3\Sigma_g^+$-state dynamics at low $\left\langle v\right\rangle$. This result is reproduced by our assumption
of vibrational relaxation and does not imply additional dephasing mechanisms. Note, that the initial distribution of $v$-state populations in the $a$-state is very different from that in the $1^3\Sigma_g^+$-state. While in the $1^3\Sigma_g^+$-state several $v$-levels are populated with similar intensities, in the $a$-state the population is peaked at $v=0$ and higher $v$-levels are much less populated by RISRS. Therefore, the $\gamma_\mathrm{D}^a$ values to good approximation reflect dephasing rates between individual $v$-levels, since the beat signal of $n$-th order coherence is mainly composed of just one beat frequency.

As we will see later, in the present case of Rb$_2$ coupled to helium nanodroplets, vibrational relaxation is likely to be the main source of dephasing. Yet, pure dephasing without population transfer does contribute to some extent. In a more complete description both dissipative as well as additional pure dephasing terms should be included to account for the observed $v$- and $\Delta v$-dependences. For the sake of restricting the model to the essential features of the problem, however, in the following discussion we focus on the model calculations that are based on vibrational relaxation.

\subsection{Numerical simulation}
\begin{figure*}
	\centering
		\includegraphics[width=\textwidth]{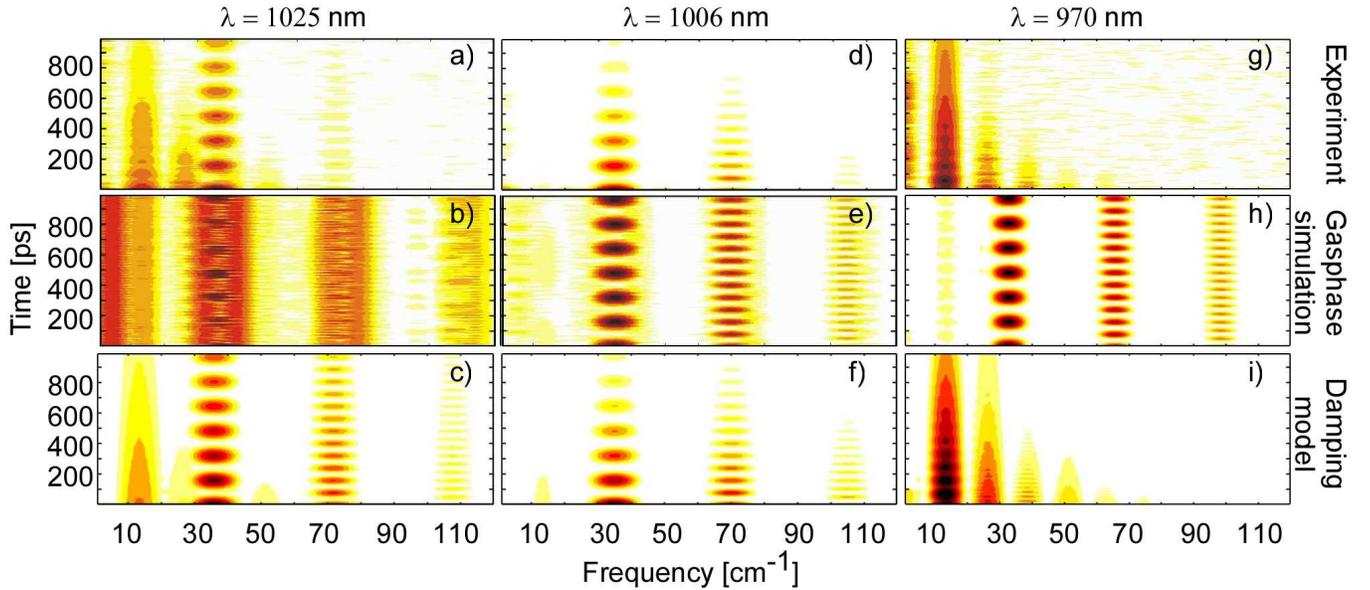}
	\caption{(Color online) Comparison between experimental (top row) and theoretical (middle and bottom row) data in spectrogram representation at selected laser wave lengths; the middle row shows the simulation of undamped vibration without coupling to the helium droplets; the bottom row shows the simulation including damping.}
	\label{fig:ExpSim}
\end{figure*}
In order to obtain a more quantitative description of the observed dynamics, the experimental data are modeled using the method outlined in Sec.~\ref{sec:simulation}. The only adjustable parameters entering the simulation are the energy relaxation rate constants in the triplet ground and first excited states, $\gamma_a$ and $\gamma_{\Sigma g}$, respectively, as well as relaxation constants for the two probe states $3^3\Pi_u$ and  $4^3\Sigma_u^+$. The resulting spectrograms of the best fits to the experimental data are displayed in Fig.~\ref{fig:ExpSim} (bottom row) for the selected laser wave lengths $\lambda=1025,$ $1006,$ and $970\,$nm. For comparison, the top row depicts the experimental data and the middle row shows the simulation when relaxation is absent ($\gamma_a=\gamma_{\Sigma g}=0$).

The transient recorded at $\lambda =1006\,$nm (center column in Fig.~\ref{fig:ExpSim}), already shown in Fig.~\ref{fig:Spectrogram1006nm}, is dominated by the fundamental as well as by the first and second overtone beats of the $1^3\Sigma_g^+$-state. The experimental data (Fig.~\ref{fig:ExpSim} (d)) are very well reproduced by the numerical simulation for a damping constant $\gamma_{\Sigma g}=0.45\,$ns$^{-1}$ (Fig.~\ref{fig:ExpSim} (f)), whereas the agreement is clearly worse when no damping is assumed (Fig.~\ref{fig:ExpSim} (e)). At laser wave lengths $\lambda = 1025\,$nm and $\lambda = 970\,$nm, WP oscillations in both ground $a^3\Sigma_u^+$ and excited states $1^3\Sigma_g^+$ are present. At $\lambda = 1025\,$nm, the excited state-dynamics clearly fades away more slowly than at $\lambda=1006\,$nm, which is in agreement with the simulated data when setting $\gamma_{\Sigma g}$=$0.36\,$ns$^{-1}$  (Fig.~\ref{fig:ExpSim} (c)). In contrast to the $1^3\Sigma_g^+$-state WP-dynamics, the $a$-state beats feature less visible dispersion and recurrences of the WP motion. This is due to the fact that predominantly the vibrational groundstate $v=0$ is populated by RISRS. Consequently, the fundamental spectral component $\omega_a\approx 13\,$cm$^{-1}$ is mainly composed of the beat frequency $(E_{v=1}-E_{v=0})/(h\,c)$, with little contributions of $(E_{v=2}-E_{v=1})/(h\,c)$ and higher level beats~\cite{Mudrich:2009}. Best agreement with the experimental data is obtained for $\gamma_a=3\,$ns$^{-1}$ (Fig.~\ref{fig:ExpSim} (c)). When no damping is assumed, the simulation clearly severely deviates from the experimental data (Fig.~\ref{fig:ExpSim} (b)). We attribute the additional spectral features to the dynamics in the higher-lying electronic state $3^3\Pi_u$ which has a similarly shaped potential curve as the $1^3\Sigma_g^+$-state. In all simulations including vibrational relaxation, the damping constants of the $3^3\Pi_u$ and of the $4^3\Sigma_u^+$-states are set to $0.1\,$ps$^{-1}$ to achieve fast damping of the corresponding WP dynamics, no direct WP signal related to these states is observed in the experiment. However, the inclusion of the two states is crucial in order to reproduce the experimentally observed WP signals in the $a$-state.
The excited state dynamics at $\lambda =970\,$nm is only visible in the time range $0$ -- $20\,$ps, after which the $a$-state dynamics prevails.  This behavior is reasonably well reproduced by the simulation when assuming very fast relaxation ($\gamma_{\Sigma g}=0.5\,$ps$^{-1}$, Fig.~\ref{fig:ExpSim} (i)). In contrast, the same simulation with $\gamma_{\Sigma g} =0$ shows a dominant contribution of the excited $1^3\Sigma_g^+$-state dynamics (Fig.~\ref{fig:ExpSim} (h)). At $\lambda =980\,$nm (not shown in Fig.~\ref{fig:ExpSim}), $1^3\Sigma_g^+$-state components are still visible during delay times $0$ -- $100\,$ps, which implies fast relaxation at a rate $\gamma_{\Sigma g}=0.01\,$ps$^{-1}$.

\begin{figure}
	\centering
		\includegraphics[width=1.0\columnwidth]{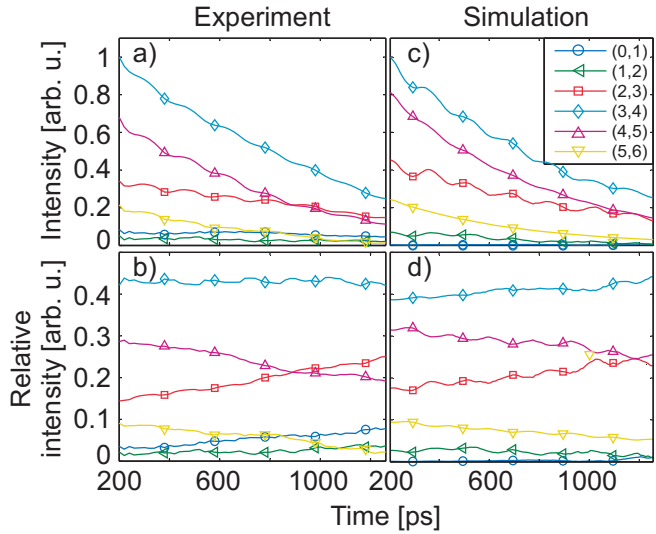}
	\caption{(Color online) Time evolution of individual beats between adjacent vibrational levels ($\Delta v=1$) of the $1^3\Sigma_g^+$-state extracted from spectrogram analysis of the transient at $\lambda=1025\,$nm with a $400\,$ps-time window ((a) and (b)) in comparison with simulated density matrix elements $|\rho_{vv+1}|$ ((c) and (d)). Plots (b) and (d) show the same data as (a) and (b) where the beats are normalized to the sum of all beats at each time step.}
	\label{fig:beats}
\end{figure}
An even more detailed verification of the numerical model is achieved by comparing the experimental and theoretical data in the spectrogram representation using a long time window of width $400\,$ps, as shown in Fig.~\ref{fig:Spectrogram1006nm} (b). The high spectral resolution retained in this analysis allows to compare the time evolution of individual beats between adjacent vibrational states. The amplitudes of individual frequency components are extracted from vertical cuts through the spectrograms at maximum positions and are plotted in Fig.~\ref{fig:beats} for $\lambda = 1025\,$nm. Panels Fig.~\ref{fig:beats} (a) and (b) represent the experimental data, where in (b) each amplitude component is normalized to the sum of all contributing beat amplitudes. Although all of the frequency components except the lowest one ($v=0$, $v=1$) decay in absolute amplitude (Fig.~\ref{fig:beats} (a)), the relative amplitudes only decrease in the case of the high-lying level beats ($4$, $5$) and ($5$, $6$), whereas the lower beats ($3$, $4$) remain constant or even rise [($0$, $1$), ($1$, $2$), and ($2$, $3$)] in amplitude in proportion to the sum of all.

The numerical simulations (Fig.~\ref{fig:beats} (c) and (d)) show the evolution of the first-order coherences of the density matrix. The good agreement justifies the assumed model based on vibrational relaxation and highlights the possibility of extracting information about the density matrix by appropriately analyzing the measured ion yields. The general decay and oscillatory behavior of individual beats is well reproduced by the numerical simulation except for the
($0$, $1$)-beat which is extraordinarily prominent in the experimental data. A slight increase of the absolute beat amplitude of the ($0$, $1$)-component can only be explained by a redistribution of population of higher-lying vibrational levels into lower-lying ones. The weak periodic modulations of both experimental and theoretical curves are reminiscent of the revival structure that becomes more pronounced as the Fourier time window is reduced.
The different decay rates for the individual vibrational beats have been discussed in terms of vibrational redistribution in the harmonic and anharmonic oscillators, (Sec. \ref{sec:DynamicsWithDissipation} and Fig.~\ref{fig:HOvsMorse}).

\begin{figure}
	\centering
		\includegraphics[width=1.1\columnwidth]{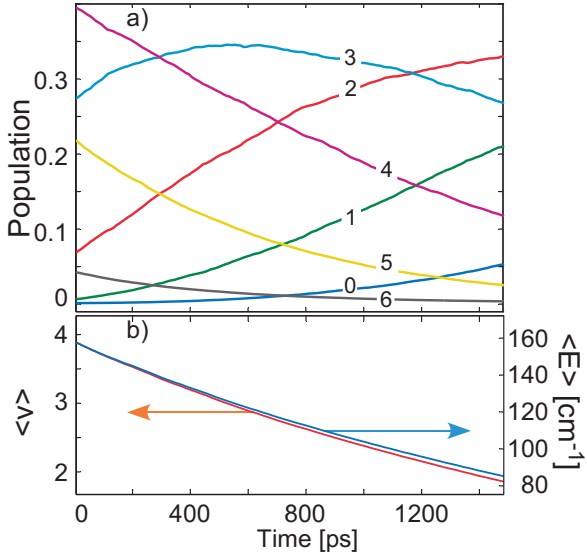}
	\caption{(Color online) (a) Time evolution of the populations of individual vibrational states extracted from the simulation at $\lambda=1025\,$nm. (b) Evolution of the vibrational level $v$ populated on average and of the mean vibrational energy $\left\langle E_v\right\rangle$.}
	\label{fig:Populations}
\end{figure}
Since all of the beats are subject to dephasing the vibrational redistribution is masked by an overall decay in the representation of absolute amplitudes in Fig.~\ref{fig:Spectrogram1006nm} (a) and (c). However, from the simulation we can extract information about the evolution of populations of the individual vibrational states. To this end, the diagonal elements of the density matrix are computed for each time step by projecting the wave function onto the vibrational eigenfunctions. The resulting populations of levels $v=0$ -- $6$ are depicted in Fig.~\ref{fig:Populations} (a). Fig.~\ref{fig:Populations} (b) shows the time evolution of the quantum number of the vibrational state that is populated on average as well as the corresponding average vibrational energy. Accordingly, at $\lambda =1025\,$nm the vibrational populations relax down by about 1.8 vibrational quanta during 1.5\,ns. The corresponding vibrational energy is reduced by about $E_{\mathrm{diss}}=73\,$cm$^{-1}$. Note that at shorter wave lengths the amount of deposited vibrational energy into the droplets in this time interval is considerably larger, \textit{e.\,g.} $E_{\mathrm{diss}}=157\,$cm$^{-1}$ at $\lambda=1006\,$nm and $E_{\mathrm{diss}}=656\,$cm$^{-1}$ at $\lambda=970\,$nm.

At such high rates of energy transfer to the helium droplets one has to consider the droplet response in terms of heating, superfluidity and cooling by evaporation of helium atoms. Dissipation of vibrational energy up to $E_{\mathrm{diss}}=656\,$cm$^{-1}$ into the droplets leads to a significant rise in droplet temperature, which may locally exceed the transition temperature to the superfluid phase ($2.17\,$K for bulk helium) of even the boiling point.
Note, however, that cooling of the droplets due to evaporation of helium atoms may counteract the heating process. In the considered energy range, effective cooling is expected to set in on a time scale of $\sim 100\,$ps~\cite{Brink:1990}. Thus, slow energy transfer from the molecules to the droplets at excitations to low-lying $v$-levels could be partly compensated by evaporation of helium atoms ($E_{\mathrm{diss}}\approx -5\,$cm$^{-1}$ per evaporated atom), whereas fast energy transfer may lead to effective heating, to subsequent local disequilibrium states and even to the breakdown of superfluidity.

However, after the time of flight of the droplets to the beam depletion detector ($\sim 1\,$ms) the complete vibrational excitation energy will be transferred to the droplets leading to massive evaporation of up to 150 helium atoms. Thus, the observed beam depletion signal could be due to the deviation of the evaporating droplets out of the beam axis instead of being the result of desorption of Rb$_2$ from the droplets, as assumed previously. This assumption is supported by the observed significant blue-shift of the beam depletion signal with respect to the pump-probe-photoionization spectrum (Fig.~11 in \cite{Mudrich:2009}). In the spectral range $\lambda \gtrsim 1000\,$nm, in which a high photoionization yield is observed but only slow dissipation, beam depletion is low.

\begin{figure}
	\centering
		\includegraphics[width=1.0\columnwidth]{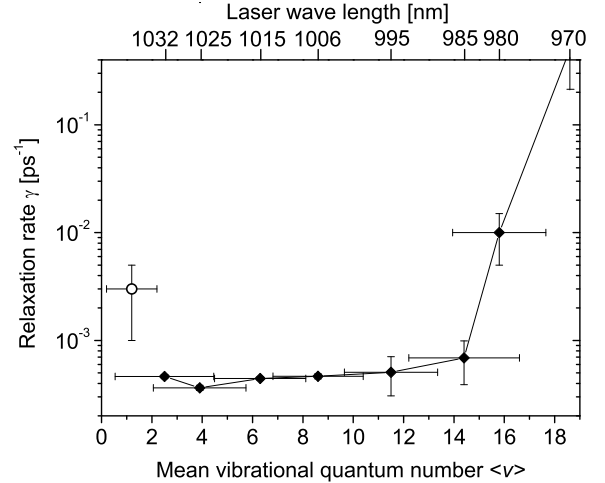}
	\caption{(Color online) Dependence of the damping parameter obtained from fitting the simulation to the experimental data as a function of the average vibrational level populated at different laser wave lengths. Filled and open symbols refer to the $1^3\Sigma_g^+$- and the $a$-states, respectively.}
	\label{fig:gamma}
\end{figure}
Having established the observed decay of vibrational beats in terms of relaxation-induced dephasing using our numerical simulation, let us finally discuss the relaxation time-constants $\gamma_i$ obtained by fitting the model to the experimental data. Fig.~\ref{fig:gamma} shows $\gamma_a$ and $\gamma_{\Sigma g}$ as a function of the average vibrational quantum number corresponding to WPs created at different laser wave lengths (top scale). Interestingly, in the range $v=2$ -- $14$ the damping parameter remains nearly constant, $\gamma_{\Sigma g}\approx 0.5\,$ns$^{-1}$, even though significantly varying decay rates $\gamma_{\mathrm{D}}$ have been measured (Fig.~\ref{fig:DecayConstants}). This discrepancy reflects the scaling behavior of dephasing times with $v$, as discussed in Sec.~\ref{sec:DynamicsWithDissipation}. The nearly $v$-independent values of $\gamma_{\Sigma g}$ are quite unexpected considering various model predictions of strongly $v$-dependent relaxation rates~\cite{Nitzan:1975,Egorov:1996,Bodo:2006}. At higher vibrational excitations $v\gtrsim 15$, though, the experimental data can only be modeled when assuming drastically increased values of $\gamma_{\Sigma g}$. The ground state relaxation rate $\gamma_a$ (open symbol in Fig.~\ref{fig:gamma}) is found to be higher by about a factor 6 as compared to the excited state rate $\gamma_{\Sigma g}$.

The nearly constant low values of $\gamma_{\Sigma g}$ in the range of small values of $v$ may be related to significant mismatch between energy spacings ($\sim 35\,$cm$^{-1}$) and excitation energies of the helium bath modes. The ripplon energies are in the range $0.1\,$cm$^{-1}$, phonon modes have energies $\sim 1\,$cm$^{-1}$, and the roton energy is about $10\,$cm$^{-1}$~\cite{Brink:1990,Chin:1995}. Consequently, the ground state vibration with lower level spacing ($\sim 13\,$cm$^{-1}$) more efficiently couples to the helium environment, which would explain the higher relaxation rate $\gamma_a$. The sharp rise of $\gamma_{\Sigma g}$ at $\left\langle v\right\rangle\gtsim 15$ could be related to the breakdown of superfluidity or even to the effect of a liquid to gas phase transition at the Rb$_2$-He interface induced by fast heating. New coupling channels, \textit{e.\,g.} the excitation of collective modes of the helium droplets that may be related to their superfluid character (\textit{i.\,e.} rotons) may also be at the origin of increasing relaxation rates~\cite{Schlesinger:2010}. At this stage, however, this assumption seems unlikely, given the fact that the vibrational energy quanta ($\sim 35\,$cm$^{-1}$) largely exceed the elementary excitations of superfluid helium droplets. Possibly, intra-molecular couplings or more complex excitation pathways leading to the ionic continuum may also be involved.


\section{Conclusion}
In conclusion, femtosecond pump-probe measurements of the vibrational wave packet dynamics of Rb$_2$ molecules attached to helium nanodroplets are analyzed using dissipative quantum simulations. In contrast to earlier interpretations, Rb$_2$ excited to triplet states are found to remain attached to the helium droplets on the time scale of the pump-probe experiments and reveal slow damping of the vibrational wave packet signal due to the interaction with the helium droplet environment. The weak system-bath coupling results in slow damping dynamics compared to the periods of vibration, which is prototypical for the applied master equation. Thus, helium droplets provide a versatile test bed for studying relaxation dynamics and cooling processes induced by a highly quantum environment.

From the detailed comparison of the experimental data in the time- and frequency domains with model calculations it is possible to deduce the evolution of the density matrix describing the vibrating Rb$_2$ molecules. While rotational-vibrational coherences as well as pure dephasing of the vibrational wave packet dynamics may play a minor role, good agreement is achieved under the model assumption of vibrational relaxation-induced dephasing. We extract damping constants for the vibrational relaxation in the lowest triplet state $a^3\Sigma_u^+$ and in the first excited state $1^3\Sigma_g^+$, $\gamma_{a}\approx 3\,$ns$^{-1}$ and $\gamma_{\Sigma g}\approx 0.5\,$ns$^{-1}$, respectively. The pronounced dependence of $\gamma_{\Sigma g}$ on the vibrational quantum number $v$  may be related to the interplay of effective heating of the droplets by fast energy transfer from the molecules and cooling due to evaporation of helium atoms. High heating rates may induce phase transitions in the droplets that affect the dynamics of attached molecules.

 Further experiments as well as modeling of the response of the helium droplets to vibronic excitations of embedded atoms and molecules are needed. In particular the dependence of dephasing and relaxation dynamics of the vibrational state quantum number may provide detailed information about the solute-solvant coupling mechanisms and may contribute to interpreting the shapes of spectral lines. Moreover, the dynamics of the desorption process of alkali atoms and molecules off the droplet surface in dependence of the atomic species and the particular vibronic state excited is still mostly unresolved.

\begin{acknowledgments}
We thank G. Stock, F. Mintert, and B. v. Issendorff for valuable discussions. Support by the Deutsche Forschungsgemeinschaft (DFG) is gratefully acknowledged. Computing resources have been provided by the Zentrum für Informationsdienste und Hochleistungsrechnen (ZIH) at the TU Dresden. M. S. is a member of the IMPRS Dresden.
\end{acknowledgments}




\end{document}